\documentclass[11pt]{article}
\usepackage{graphicx}
\usepackage{amsfonts}
\usepackage{amsmath}
\usepackage{amssymb}
\usepackage{latexsym}
\usepackage{color}
\usepackage{colordvi}
\usepackage{epsfig}
\usepackage{array}
\usepackage{pifont}
\usepackage{axodraw}
\usepackage{citesort}
\usepackage{braket}
\newcommand{\msbar}{\overline{\mbox{{\sc ms}}}}
\newcommand{\VA}{\braket{A^2}}

\newcommand{\beq}{\begin{eqnarray}}
\newcommand{\eeq}{\end{eqnarray}}

\newcommand{\be}{\begin{equation}}
\newcommand{\ee}{\end{equation}}
\newcommand{\lwrsim}{\raise0.3ex\hbox{$<$\kern-0.75em\raise-1.1ex\hbox{$\sim$}}}

\def\C2#1#2{({\cal C}_2)_{#1}^{#2}}

\def\eq#1{Eq.~(\ref{#1})}
\def\altura#1{\rule[0cm]{0cm}{#1cm}}

\newcommand{\ghThreeOneR}{\begin{picture}(150,45)(0,0)
\SetWidth{1.2}
\DashArrowLine(12.5,0)(37.5,0){5}
\DashArrowLine(37.5,0)(112.5,0){5}
\DashArrowLine(112.5,0)(137.5,0){5}
\SetWidth{1}
\Vertex(37.5,0){2}
\Vertex(112.5,0){2}
\Vertex(40,31){2}
\Gluon(37.5,0)(37.5,48){-4}{4}
\GlueArc(67.5,0)(45,70,135){-4}{4}
\GlueArc(67.5,0)(45,0,70){-4}{5}
\CCirc(78,40){10}{Black}{Blue}
\Text(135,5)[]{q}
\Text(20,5)[]{$\varepsilon$}
\Text(25,47)[]{q-$\varepsilon$}
\end{picture}}
\newcommand{\ghThreeOneRS}{\begin{picture}(150,45)(0,0)
\SetWidth{1.2}
\DashArrowLine(12.5,0)(37.5,0){5}
\DashArrowLine(37.5,0)(112.5,0){5}
\DashArrowLine(112.5,0)(137.5,0){5}
\SetWidth{1}
\Vertex(37.5,0){2}
\Vertex(112.5,0){2}
\Vertex(40,31){2}
\Gluon(37.5,0)(37.5,48){-4}{4}
\GlueArc(67.5,0)(45,70,135){-4}{4}
\GlueArc(67.5,0)(45,0,70){-4}{5}
\CCirc(78,40){10}{Black}{Blue}
\Text(135,5)[]{q}
\Text(20,5)[]{k}
\Text(25,47)[]{q-k}
\end{picture}}
\newcommand{\ghThreeOneRSs}{\begin{picture}(150,45)(0,0)
\SetWidth{1.2}
\DashArrowLine(12.5,0)(37.5,0){5}
\DashArrowLine(37.5,0)(112.5,0){5}
\DashArrowLine(112.5,0)(137.5,0){5}
\SetWidth{1}
\Vertex(37.5,0){2}
\Vertex(112.5,0){2}
\Vertex(109,34){2}
\Gluon(112.5,0)(112.5,48){4}{4}
\GlueArc(82.5,0)(45,53,100){-4}{3}
\GlueArc(82.5,0)(45,100,180){-4}{5}
\CCirc(70,40){10}{Black}{Blue}
\Text(135,5)[]{q}
\Text(20,5)[]{k}
\Text(125,47)[]{q-k}
\end{picture}}
\newcommand{\ghThreeTwo}{\begin{picture}(120,40)(0,0)
\SetWidth{1.2}
\DashArrowLine(10,20)(30,20){5}
\DashArrowLine(30,20)(90,20){5}
\DashArrowLine(90,20)(110,20){5}
\SetWidth{1}
\Vertex(30,20){2}
\Vertex(90,20){2}
\Gluon(60,20)(60,-10){4}{3}
\GlueArc(60,20)(30,0,75){-4}{4}
\GlueArc(60,20)(30,105,180){-4}{4}
\CCirc(60,50){10}{Black}{Blue}
\end{picture}}
\newcommand{\ghThreeThree}{\begin{picture}(130,25)(0,0)
\SetWidth{1.2}
\DashArrowLine(10,0)(30,0){5}
\DashArrowLine(30,0)(90,0){5}
\DashArrowLine(90,0)(120,0){5}
\SetWidth{1}
\Vertex(30,0){2}
\Vertex(90,0){2}
\Vertex(112,0){2}
\Gluon(112,0)(112,40){4}{4}
\GlueArc(60,0)(30,0,75){-4}{4}
\GlueArc(60,0)(30,105,180){-4}{4}
\CCirc(60,30){10}{Black}{Blue}
\end{picture}}

\newcommand{\ghThreeThreeRS}{\begin{picture}(130,25)(0,0)
\SetWidth{1.2}
\DashArrowLine(10,0)(30,0){5}
\DashArrowLine(30,0)(90,0){5}
\DashArrowLine(90,0)(120,0){5}
\SetWidth{1}
\Vertex(30,0){2}
\Vertex(90,0){2}
\Vertex(112,0){2}
\Gluon(30,0)(30,40){4}{4}
\GlueArc(82,0)(30,0,75){-4}{4}
\GlueArc(82,0)(30,105,180){-4}{4}
\CCirc(82,30){10}{Black}{Blue}
\Text(10,7)[]{k}
\Text(15,35)[]{q-k}
\Text(125,7)[]{q}
\end{picture}}
\newcommand{\ghThreeFour}{\begin{picture}(120,40)(0,0)
\SetWidth{1.2}
\DashArrowLine(110,-10)(110,20){5}
\DashArrowLine(110,20)(110,50){5}
\SetWidth{1}
\Gluon(10,20)(30,20){4}{2}
\Gluon(30,20)(90,20){4}{6}
\Gluon(90,20)(110,20){4}{2}
\Vertex(30,20){2}
\Vertex(90,20){2}
\Vertex(110,20){2}
\GlueArc(60,20)(30,0,75){-4}{4}
\GlueArc(60,20)(30,105,180){-4}{4}
\CCirc(60,50){10}{Black}{Blue}
\end{picture}}

\newcommand{\ghThreeFive}{\begin{picture}(120,40)(0,0)
\SetWidth{1.2}
\DashArrowLine(110,-10)(110,10){5}
\DashArrowLine(110,10)(110,50){5}
\SetWidth{1}
\Gluon(10,10)(60,10){4}{6}
\Gluon(60,10)(110,10){4}{6}
\Vertex(60,10){2}
\Vertex(110,10){2}
\GlueArc(60,35)(25,105,270){4}{7}
\GlueArc(60,35)(25,-90,75){4}{7}
\CCirc(60,60){10}{Black}{Blue}
\end{picture}}

\title{\textbf{On the leading OPE corrections to the ghost-gluon vertex and
the Taylor theorem}}

\author{Ph. Boucaud$^{1}$,  D. Dudal$^{2}$, J.P. Leroy$^{1}$, O. P\`ene$^{1}$, J. Rodr\'{\i}guez-Quintero$^{3}$}

\date{}

\begin{document} 

\maketitle

\begin{center}

$^{1}$ Laboratoire de Physique Th\'eorique, \\
Universit\'e de Paris XI; B\^atiment 211, 91405 Orsay Cedex, France\\
$^{2}$ Department of Physics and Astronomy, Ghent University, Krijgslaan 281 S9 9000 Gent, Belgium\\
$^{3}$ Dpto. F\'isica Aplicada, Fac. Ciencias Experimentales,\\
Universidad de Huelva, 21071 Huelva, Spain
\end{center}

\begin{abstract}

This brief note is devoted to a study of genuine non-perturbative corrections
to the Landau gauge ghost-gluon vertex in terms of the non-vanishing dimension-two
gluon condensate. We prove these corrections to give account of current SU(2) lattice data
for the vertex with different kinematical configurations in the domain of intermediate
momenta, roughly above 2-3 GeV. We pay special attention to the kinematical
limit which the bare vertex takes for its tree-level expression at any perturbative
order, according to the well-known Taylor theorem.  Based on our OPE analysis, we also present
a simple model for the vertex, in acceptable agreement with the lattice data also in the
IR domain.

\end{abstract}

\vspace{-14cm}
\begin{flushright}
{\small UHU-FP/11-025}\\
{\small LPT-Orsay/11-72}\\
\end{flushright}
\vspace{14cm}




\section{Introduction}

The infrared properties of the Landau-gauge QCD Green functions have motivated many studies
in the last few years, mainly involving both
lattice (see for instance Refs.~\cite{Cucchieri:2007rg,Cucchieri:2008fc,Cucchieri:2009zt,Costa:2010pp,Oliveira:2010xc,Bogolubsky:2007bw,Bogolubsky:2009dc,Bornyakov:2009ug})
and continuum approaches (see for instance Refs.~\cite{Aguilar:2008xm,Binosi:2009qm,Boucaud:2008ky,Boucaud:2008ji,Boucaud:2010gr,RodriguezQuintero:2010wy,Fischer:2008uz,Fischer:2009tn,Alkofer:2008jy} using Dyson-Schwinger equations (DSE), \cite{Dudal:2005na,Dudal:2007cw,Dudal:2008sp,Dudal:2010tf} using the refined Gribov-Zwanziger formalism or \cite{Tissier:2010ts,Tissier:2011ey} using the Curci-Ferrari model as an effective description).
Most of the DSE analysis take advantage of approximating the ghost-gluon vertex by a constant when truncating
the infinite tower of the relevant equations. To be more precise, only the behaviour of the involved transverse
form factor needs to be approximated by a constant for the purpose of truncating the ghost
propagator DSE (see also the recent paper \cite{Pennington:2011xs}).
However, this ghost propagator DSE (GPDSE) appears to be a cornerstone for the analysis of
infrared DSE solutions for the gluon correlations functions: the two classes of solutions, namely ``decoupling''
and ``scaling'', depend on the regular or singular behaviour of the ghost dressing function at zero-momentum
and on how this constrains the gluon behaviour via the GPDSE~\cite{Boucaud:2008ky}.
Most of the ammo for the approximation of the ghost-gluon vertex and GPDSE truncation is mainly provided by
the Taylor theorem which is widely known as a non-renormalization one~\cite{Taylor:1971ff}.
This claims that, in the particular kinematical configurations defined by a vanishing incoming ghost-momentum,
no non-zero radiative correction survives for the Landau-gauge ghost-gluon vertex, which takes thus its tree-level
expression at any perturbative order~\cite{Taylor:1971ff}. This statement entails two important consequences: that (i)
the bare ghost-gluon vertex is UV-finite for any kinematical configuration,
and that (ii) in the specific MOM scheme where the renormalization point is taken with a vanishing
incoming ghost momentum (which we named the Taylor scheme --T-scheme-- in \cite{Boucaud:2008gn})
the renormalization constant for
the ghost-gluon vertex\footnote{It  can also be  straightforwardly concluded that the ghost-gluon vertex
renormalization constant, $\widetilde{Z}_1$, is exactly 1 in the $\msbar$ scheme.}  is exactly~1.
Then, one can use  those two  conclusions to write (schematically) the renormalized GPDSE as
\beq\label{GPDSEsq}
\frac 1 {F(k^2)} \ = \ 1 \ + \ \alpha_T(\mu^2) \ \int d^4q \ K(k,q) H_1^{\rm bare}(q,k) F(q^2) \ ,
\eeq
where $\mu$ is the subtraction momentum for all the renormalized quantities, $F$ is the ghost-propagator dressing
function, $K(k,q)$ is the part of the kernel which does not involve the ghost-gluon vertex, $\alpha_T$ is the running coupling in the
T-scheme\footnote{The coupling in T-scheme has been profusely computed on the lattice and confronted to perturbative predictions to
estimate $\Lambda_{\overline{\rm MS}}$~\cite{Boucaud:2008gn,Blossier:2010ky,Sternbeck:2010xu}. It has been also used to
define an effective charge that could be applied for phenomenological purposes~\cite{Aguilar:2009nf}.},
and $H_1^{\rm bare}$ is the transverse form factor of the finite bare ghost-gluon vertex,
defined by
\beq\label{vertMinko}
\Gamma^{abc}_{\mu}(-q,k;q-k) \ = \ - g_0 f^{abc}  \
\left( \altura{0.5} q_\mu H_1(q,k) + (q-k)_\mu H_2(q,k) \right) \ .
\eeq
The kernel $K$ corresponds to the gluon propagator, its transverse character in the Landau gauge allows to project out the $H_2$-form factor from
expression \eqref{vertMinko}. Therefore, we have chosen to call the surviving piece, i.e.~$H_1$, the transverse form factor. Extending this
nomenclature, we call $H_2$ the longitudinal vertex form factor. The kernel $K$ can be either built by invoking lattice data (as done,
for instance, in ref.~\cite{Boucaud:2008ji})
or determined by solving the coupled system of \eq{GPDSEsq} and the gluon propagator DSE. In both cases,
a precise determination of the ghost dressing function from \eq{GPDSEsq},
to be confronted for instance with lattice data, requires a correspondingly precise knowledge of the transverse form factor $H_1$.
Thus, the ghost-gluon vertex plays an important quantitative and qualitative
r\^ole for the infrared DSE analysis of Yang-Mills QCD Green functions. In fact, with the usual bare ghost-gluon vertex approximation,
namely that the bare transverse form factor is 1 over all the momentum range,
it is readily appreciated from e.g.~the results of \cite{Boucaud:2008ji,Boucaud:2011ug} that when the lattice data for the gluon propagator is fed
into the GPDSE, there is a qualitative difference with the ghost propagator lattice data, unless one boosts by hand the input for the coupling constant
at the renormalization point to an artificially large value. Another way-out~\cite{Boucaud:2008ji}
is to impose the bare transverse form factor to take a constant,
clearly different from 1, to borrow the boost of the coupling input needed to describe lattice data\footnote{As it is shown in
refs.~\cite{Pp:TNT11,Dudal:TNT11}, taking a constant larger than 1 for the bare $H_1$ helps to account better, but not enough,
for the ghost propagator lattice data with the results of the integration of the GPDSE. Some non-perturbative structure
for the ghost-gluon vertex is thus needed, and it is also preliminary shown there that the sort of inputs we can obtain with the OPE analysis in
this work might very well lead to a good description of lattice data.}.
This is strongly hinting towards a missing piece in the ghost-gluon vertex, which, if stronger than a tree level one,
will lift the output of the GPDSE up to the ghost lattice data (see refs.~\cite{Pp:TNT11,Dudal:TNT11}).
Of course, before this can be precisely tested in practice, we need to cook up a reliable
input for the ghost-gluon vertex. How to do this will be investigated in this work.\\

On the other hand, the study of the running of the QCD Green functions triggered some lattice
works~\cite{Alles:1996ka,Becirevic:1999uc,Becirevic:1999hj,Boucaud:2000ey,Boucaud:2000nd,Boucaud:2001qz,Boucaud:2001st,DeSoto:2001qx,Boucaud:2005xn,Bonnet:2001uh,Bonnet:2000kw,Williams:1999xk},
many of them aiming to determining the $\Lambda_{\rm QCD}$ parameter, which concluded that the  impact of the
non-perturbative corrections coming from the dimension-two gluon condensate $\VA$ was not
negligible~\cite{Boucaud:2000nd,Boucaud:2001qz,Boucaud:2001st,DeSoto:2001qx,Boucaud:2005xn}.
In particular, in ref.~\cite{Boucaud:2008gn} we computed the running coupling renormalized in the Taylor scheme from
the lattice in pure Yang-Mills gauge theory and proved the necessity of including  non-perturbative power corrections to describe properly
the running over a momenta window from around 3 to 5 GeV. In the T-scheme, the configuration of momenta for the ghost-gluon
vertex is the one ``protected'' by the non-renormalization theorem and, at least in perturbation theory to any order, no
correction to the bare ghost-gluon vertex is expected beyond the tree-level expression. However, non-perturbative
OPE corrections should be expected for any other kinematical configuration and, furthermore, it is interesting
to examine whether a genuine non-perturbative correction might survive in the particular configuration where
Taylor's argument works. More interesting will be to investigate whether, for the same price,
these non-perturbative corrections related to $\VA$ might suggest, apart from describing the current lattice data
inside the appropriate momenta window, a reliable input for the ghost-gluon vertex we mentioned above.
This is the main purpose of this note.

\section{The OPE for the ghost-gluon vertex}
\label{sec:OPEsym}

The OPE procedure for the Landau-gauge ghost-gluon vertex,
\beq
V^{abc}_\mu(-q,k;q-k) &=&
\Gamma^{a'b'c'}_{\mu'}(-q,k;q-k) \ G^{bb'}_{\mu\mu'}(q-k) \ F^{aa'}(q) \
F^{cc'}(k)
\nonumber \\
&=&
\int d^4y \ d^4x \ e^{i (q-k) \cdot x} \ e^{i k \cdot y}
\Braket{ \ T\left( c^c(y) A^b_\mu(x) \overline{c}^a(0) \right)}
\eeq
is quite similar to the one described in ref.~\cite{Boucaud:2008gn}.
Here, the OPE expansion shall read

\beq\label{OPEvertexS}
V^{abc}_\mu(-q,k;q-k) &=&
\left(d_0\right)^{abc}_{\mu}(q,k) \
\nonumber \\
&+&
\left(d_2\right)^{abc\mu'\nu'}_{\mu a'b'}(q,k) \
\Braket{ \  :  A_{\mu'}^{a'}(0) A_{\nu'}^{b'}(0) :}
\ + \ \cdots
\eeq
where $d_0$ accounts for the purely perturbative contribution
to the vertex, while
\beq
w^{a b c}_{\mu} &=&
\left(d_2\right)^{abc\mu'\nu'}_{\mu a'b'}(q,k) \ \delta^{a'b'} g_{\mu'\nu'}
\nonumber \\
&=& 2 I^{[1]} + 2 I_s^{[1]} + 2 I^{[2]} + 4 I^{[3]} +2I^{[4]} + I^{[5]}
\eeq
where\footnote{$N_C$ is the number of colours.}
\beq\label{I1S}
I^{[1]} & = & \ghThreeOneRS \nonumber
\\
%
&=&   \rule[0cm]{0cm}{0.9cm}
\frac{N_C}{2} \ g^2 \
\frac{(q-k)_\sigma q_\sigma}{q^2 (q-k)^2} \
V^{abc}_{{\rm tree}, \mu}(-q,k;q-k) \ ,
\eeq
the tree-level ghost-gluon vertex being
\beq
V^{abc}_{{\rm tree}, \mu}(-q,k;q-k) \ = \ - i \frac{g}{k^2 q^2 (q-k)^2} \ f^{abc} \
q_{\mu_2} \ g^{\perp}_{\mu_2 \mu}(q-k) \ ,
\eeq
while
\beq\label{I1vertexSs}
I_s^{[1]} &=&
\ghThreeOneRSs \ = \
I^{[1]}
\left\{
\begin{array}{c}
q \to -k
\\ k \to -q
\end{array}
\right\} \nonumber  \\
&=&
\altura{0.9}
- \frac{N_C}{2} \ g^2 \
\frac{(q-k)_\sigma k_\sigma}{k^2 (q-k)^2} \
V^{abc}_{{\rm tree}, \mu}(-q,k;q-k) \,,
\eeq
and
\beq\label{I2S}
I^{[2]} \altura{1.8} & = &  \ghThreeTwo \nonumber  \\
&=&   \altura{0.8}
\frac {N_C} {4} g^2 \frac {k_\sigma q_\sigma}{q^2 k^2} \ V_{{\rm tree},\mu}^{a b c}(-q,k;q-k) \  \ .
\eeq
The other contributions, $I^{[3,4,5]}$, can be immediately written using the previous results:

\beq\label{I3S}
I^{[3]} \ &=& \frac 1 2 \left( \ \ghThreeThree + \ghThreeThreeRS \ \right) \nonumber \\
&=& \left( \frac {N_C}{4} \frac{g^2}{k^2} + \ \frac {N_C} {4} \frac{g^2}{q^2} \right) V_{{\rm tree},\mu}^{a b c}(-q,k;q-k)
\eeq
and
\beq\label{I45S} \altura{1.5}
2I^{[4]} +  I^{[5]} \ &=& \ 2\times \ghThreeFour \ +  \ghThreeFive
\nonumber \\
&=&  \ N_C \frac{g^2}{(q-k)^2} \ V_{{\rm tree},\mu}^{a b c}(-q,k;q-k)
\eeq
In all cases, the blue bubble refers to a contraction of the colour and Lorentz indices with $\frac{1}{2}\delta_{ab}g_{\mu\nu}$. We have also introduced the notation $g^\perp_{\mu\nu}(\ell)=g_{\mu\nu}-\frac{\ell_\mu \ell_\nu}{\ell^2}$ for the transverse projector.\\

Then, one obtains
\beq
w^{abc}_\mu \ = \ g^2 \ \left( \altura{0.8} s_V(q,k) + s_F(k) + s_F(q) + s_G(k-q) \right)\,
V_{{\rm tree},\mu}^{a b c}(-q,k;q-k)
\eeq
with
\beq\label{eq:sVsG}
s_G(q) &=& s_F(q) \ = \frac{N_C}{q^2}\,, \nonumber \\
s_V(q,k) &=& \frac {N_C} {2} \left( 2 \ \frac{(q-k)\cdot q}{q^2 (q-k)^2} + 2 \ \frac{(k-q)\cdot k}{k^2 (q-k)^2} + \frac{k\cdot q}{k^2 q^2} \right)  \,.
\eeq
$s_F$ (resp. $s_G$)  comes from the OPE corrections to the external ghost (resp. gluon)
propagator, {\sl i.e.}  from the non-proper diagrams in Eqs.~(\ref{I3S},\ref{I45S})
and $s_V$ from the proper vertex correction in Eqs.~(\ref{I1S},\ref{I2S}).
The OPE corrections to the proper vertex can be obtained from there, remembering that
the bare ghost-gluon vertex reads as:
\beq\label{vertMinko2}
\Gamma^{abc}_{\mu}(-q,k;q-k) \ = \ - g_0 f^{abc}  \
\left( \altura{0.6} q_\mu H_1(q,k) + (q-k)_\mu H_2(q,k) \right) \ .
\eeq
Then, in Landau gauge, only the form factor $H_1$ survives in the Green function to give:
\beq
V^{abc}_{\mu}(-q,k;q-k) &=&
-i g_0 f^{abc} q_{\mu'} g^{\perp}_{\mu'\mu}(q-k) H_1(q,k) \ G((q-k)^2) \ F(q^2) \ F(k^2) \ ,
\eeq
where $G$ and $F$ are the gluon and ghost dressing functions for which we previously computed
the non-perturbative OPE corrections. Thus, one would have:
\beq\label{eq:H1OPE}
H_1(q,k) \ = \ H_1^{\rm pert}(q,k) \left( 1 + s_V(q,k) \ \frac{\braket{A^2}}{4 (N_C^2-1)}
+  \  {\cal O}(g^4,q^{-4}, k^{-4},q^{-2}k^{-2})\ \right).
\eeq
All the non-proper corrections have been removed in the usual way.

\section{A model for the Euclidean ghost-gluon vertex}

We will now devote this section to model the transverse form factor of the Euclidean ghost-gluon
vertex, $H_1$, on the ground provided by \eq{eq:H1OPE}, and compare the result with some recent
lattice data for this form factor computed in different kinematical configurations.

\subsection{The model}

In Euclidean metrics the bare ghost-gluon vertex can be written  very similarly  as:
\beq
\Gamma^{abc}_{{\rm bare},\mu}(-q,k;q-k) \ = \ i g_0 f^{abc}  \
\left( \altura{0.6} q_\mu H_1(q,k) + (q-k)_\mu H_2(q,k) \right) \,,
\eeq
where the form factor $H_1$ plays a crucial r\^ole when solving
the ghost-propagator Dyson-Schwinger equation (GPDSE) as
discussed above.
The OPE non-perturbative corrections to the form
factor $H_1$ were obtained in \eq{eq:H1OPE}, but that result
is in principle only reliable for large enough $k,q$ and $q-k$ since the  SVZ factorization
on which it relies might not be valid for low  momenta. Nevertheless, in the following we will propose
a very simple conjecture to extend \eq{eq:H1OPE} to any momenta, which will provide us with
a calculational model for the ghost-gluon vertex to continue research with.
In particular, the perturbative part of the ghost-gluon vertex is usually approximated by
a constant behaviour\footnote{At least for large momenta, this seems to be the case in
lattice simulations~\cite{Cucchieri:2008qm,Sternbeck:2005re} for several kinematical configurations, and
so is confirmed also by the perturbative calculations in Refs~\cite{Chetyrkin:2000dq,Davydychev:1996pb}.}; if we then apply a
finite renormalization prescription such that
\beq
\widetilde{Z}_1(\mu^2) \left. H_1(q,k) \right|_{\mu^2} \ = \ 1 \ ,
\eeq
where the renormalization momentum, $\mu^2$, for a given kinematical configuration
(for instance, $q-k=0$ and $q^2=k^2=\mu^2$) is chosen to be large enough,
on the basis of \eq{eq:H1OPE}, we can conjecture that
\beq\label{eq:H1model}
H_1(q^2,k^2,\theta) &=&
\widetilde{Z}_1^{-1} \ \left[ \altura{0.95} 1 \ + \
\frac{ N_C g^2 \braket{A^2}}{8 (N_C^2-1)} \right. \nonumber \\
&\times&
\left( \frac{\sqrt{k^2q^2} \cos\theta}{k^2 q^2 + m_{\rm IR}^4}
+ \ 2 \ \frac{q^2-\sqrt{k^2q^2} \cos\theta}{q^2 (q^2+k^2 - 2\sqrt{q^2k^2}\cos\theta) + m_{\rm IR}^4}
\right.
\nonumber \\
&&
\left. \left.
+ \ 2 \ \frac{k^2-\sqrt{k^2q^2} \cos\theta}{k^2 (q^2+k^2 - 2\sqrt{q^2k^2}\cos\theta) + m_{\rm IR}^4}
 \right)
\altura{0.95}
\right] \ ,
\eeq
gives a reasonable description  of the ghost-gluon form factor $H_1$ all over the range
of its momenta $q$ and $k$, where $\theta$ stands for the angle between them. The purpose of
\eq{eq:H1model} is to keep the main features of the momentum behaviour of the ghost-gluon
form factor provided by the OPE analysis and to give a rough description of the deep IR
only with the introduction of some IR mass scale, $m_{\rm IR}$, which is mainly aimed to
avoid the spurious singularities resulting from the OPE expansion in momentum inverse powers.

It should be noted that the OPE technology implies first a hierarchic expansion in terms
of local operators and next a factorization assumption that
allows for all the large-distance information to be encoded in the condensates of those
local operators in the QCD vacuum. If the momentum to be described decreases, more condensates of
higher and higher dimensions should be needed, but also the Wilson coefficient containing
the short-distance information should be computed at a higher loop level. To try to sum up
all this is, if not impossible, a daunting task. This is why the OPE is applied to provide
only an asymptotic expansion which is reliable to account for some QCD phenomenology, but nevertheless it can
also successfully reproduce quenched and unquenched lattice results for many
Green functions in some intermediate momentum region~\cite{Boucaud:2008gn,Boucaud:2001st,DeSoto:2001qx,Boucaud:2005xn}.
Then, it is important to stress here that the model vertex
\eqref{eq:H1model} will never be obtainable from the OPE in QCD, as it is just impossible to describe with the OPE 
whatever quantity over a full momentum range. Very recently, a Landau gauge version of the Curci-Ferrari model has 
been shown to give, at the first perturbative order, a reliable description of quenched QCD lattice
results~\cite{Tissier:2010ts,Tissier:2011ey}. There is no reason to think that the same approach could also 
account for the operator expansion and factorization sum rules. As a matter of fact, one would obtain that applying
the OPE technology with a massive gluon propagator, for instance with a frozen mass $M$, 
implies to replace 
\beq
\frac 1 {(q-k)^2} \ \to \ \frac 1 {(q-k)^2 + M^2}
\eeq
in Eqs.~(\ref{eq:sVsG},\ref{eq:H1OPE}) and this does not totally cure the above mentioned
spurious singularities coming from the expansion in momentum inverse powers. In fact, assuming that $M$ is 
the typical mass scale of a theory, the OPE machinery only intends to describe physics for momenta (much) 
larger than $M$. However, lattice results, as can be seen in the next subsection, and the Taylor theorem in its
particular limit prove that no singularity appear in the IR. This is why we
introduce in \eq{eq:H1model} the mass scale, $m_{\rm IR}$, that can be thought as
some sort of IR regulator that should be related to some kind of effective gluon mass scale.

The proposed expression is thus supposed to describe the vertex for the whole range of possible momenta, 
as needed to be plugged into the GPDSE in future work, while any OPE analysis, from its very definition, 
can only be valid for relatively large momenta.
The proposal \eqref{eq:H1model} has however the right asymptotic momentum behaviour, as compared to the OPE 
result \eqref{eq:H1OPE}.
Its validity for lower momenta will shortly be checked by a confrontation with the available lattice data in
the next subsection. Said otherwise, \eq{eq:H1model} can be also seen as a parametrization to describe lattice data
(to be properly plugged into the GPDSE) that, behaving asymptotically as the OPE predicts, benefits of the fact
that free parameters as $\VA$ can be borrowed from (or compared with) different and independent analyses.
It is anyhow very important to stress that it is the OPE analysis leading to \eqref{eq:H1OPE}
providing us with the kinematical structure of \eqref{eq:H1model}, at least in the
intermediate region, and that the infrared completion has nothing to do with this. Thus, the success in the
confrontation with the available lattice data, with different kinematical combinations
for the form factor, will also confirm that the OPE analysis works in the
appropriate region and it also gives an independent estimate of $\VA$.

Furthermore, to give further credibility to the vertex \eqref{eq:H1model}, we also plan a future study of the 
ghost-gluon vertex from the analytical viewpoint. For example, it would be instructive to find out if the 
refined Gribov-Zwanziger (RGZ)
computational scheme could give access to a vertex of the type modeled here, since this formalism incorporates
effects of non-perturbative dimension 2 condensates in a controllable, renormalizable setting, applicable to the
whole range of momenta. It is worth to remark that, as an enlightening indication, the gluon propagator
in the RGZ scheme can be asymptotically expanded to agree with its OPE prediction stemming from the gluon condensate, $\VA$;
however, the deep infrared structure of the propagator clearly escapes from the OPE analysis and depends, for instance,
on additional dimension two-condensates and on the Gribov-Zwanziger parameter defined through the so-called
Gribov horizon condition~\cite{Dudal:2005na,Dudal:2007cw,Dudal:2008sp,Dudal:2010tf}.
An alternative could be the above-mentioned Landau gauge version of the Curci-Ferrari model,
which was already tested at the propagator level in \cite{Tissier:2010ts,Tissier:2011ey} but that
can be also applied to give a prediction for the ghost-gluon vertex.

Of course, $\mu^2$ being large enough, $\widetilde{Z}_1$ can be
approximated by some constant value for any $\mu^2$ because the logarithmic behaviour
of the ghost-gluon in perturbation theory has been proven to be very
smooth~\cite{Chetyrkin:2000dq,Davydychev:1996pb}. Thus
\eq{eq:H1model} provides us with a very economical model because one
needs nothing but some infrared mass parameter which, being related to the gluon mass,
is expected to be $\sim 1$ GeV, to parametrize the deep infrared behaviour of
the ghost-gluon transverse form factor.
It should be noted that, although the ghost-gluon vertex form factors do not diverge, the
gluon condensate does: it is the product of the Wilson coefficient and of the condensate which
is expected to remain finite thanks to a delicate compensation of singularities,
in the sum-rules approach. Then, both the Wilson coefficient and the condensate should be renormalized,
by fixing some particular prescription, and both shall depend on a chosen renormalization momentum.
This is  explained in detail in the work of ref.~\cite{Blossier:2010vt} where the Wilson coefficient for the
quark propagator is included at  order ${\cal O}(\alpha^4)$.
Since we carry out the computation of the Wilson coefficient in Eqs.~(\ref{I1S}-\ref{I45S}) only at  tree-level
and neglect its logarithmic dependence, we do not specify any renormalization momentum
for the gluon condensate.

\subsection{Comparing with available lattice data}

\begin{figure}[h]
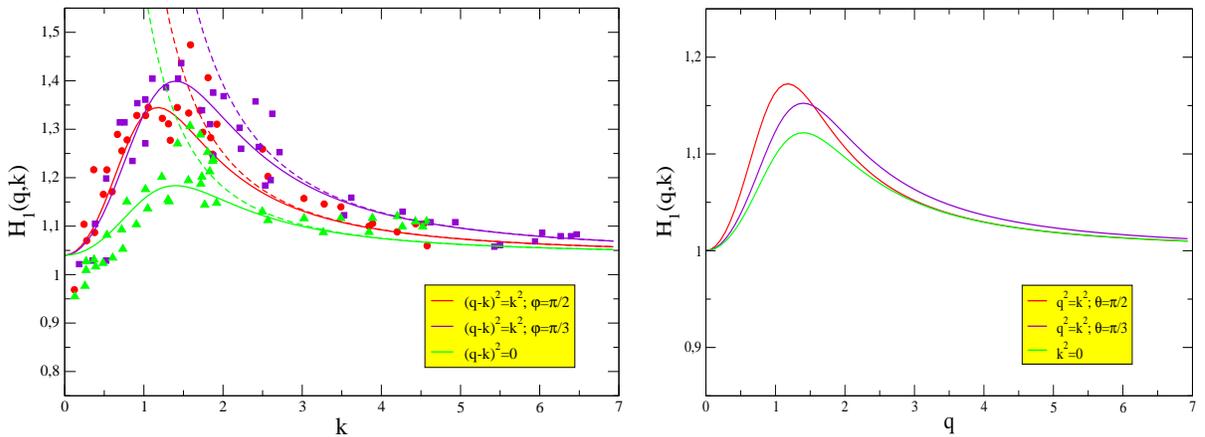

\begin{center}
\begin{tabular}{cc}
\includegraphics[width=8.2cm]{figs/H1SU2.eps} &
\includegraphics[width=7.2cm,height=5.7cm]{figs/H1SU3.eps}
\end{tabular}
\end{center}
\caption{\small (left) The predictions of the model for $H_1$ (solid lines) in eq.~(\ref{eq:H1model}) 
and the OPE result without infrared completion (dotted lines) in eq.~\eqref{eq:H1OPE} confronted
to the SU(2) lattice data borrowed from ref.~\cite{Cucchieri:2008qm} for the three kinematical
configurations described in the main text: $(k-q)^2=q^2$ for $\varphi=\pi/3$ (violet)
and $\varphi=\pi/2$ (red) and $(k-q)^2=0$ (green).
(right) The results of the model for $H_1$ in the SU(3) case and for the following kinematical configurations:
$k^2=q^2$ for $\theta=\pi/3$ (violet) and $\theta=\pi/2$ (red) and $k^2=0$ (green).}
\label{fig:lat}
\end{figure}

Some  SU(2) and SU(3) lattice results for the ghost-gluon vertex are available in the literature (see for instance
Refs.~\cite{Cucchieri:2008qm,Sternbeck:2005re}). In particular, Cucchieri et al.   have computed the tree-level-tensor
form factor in SU(2) for three kinematics  ($p \equiv q - k$ is the gluon momentum and  $\varphi$ the angle
between the gluon and ghost momenta) : (1) $p^2=q^2$ and $\varphi=\pi/2$,  (2)  $p^2=0$
and (3) $\varphi=\pi/3$ with $p^2=q^2=k^2$ (\cite{Cucchieri:2008qm}).
As can be seen in the left pane of Fig.~\ref{fig:lat},
 when we plug $g^2 \langle A^2\rangle=6$ GeV$^2$, $m_{\rm IR}=1.4$ GeV  and $\widetilde{Z}_1^{-1}=1.04$
 into \eq{eq:H1model}, the prediction for the form factor $H_1$ appears to agree pretty well
 with the SU(2) lattice data for the three kinematical configurations above mentioned. It should be also noted 
 that the OPE prediction without the infrared completion through the introduction of $m_{\rm IR}$ 
 given by Eqs.~(\ref{eq:sVsG},\ref{eq:H1OPE}), plotted with dotted lines, also accounts very well for 
 lattice data in the intermediate momenta region, above roughly $2.5$ GeV.  
 Furthermore, the
 value for the gluon condensate lies in the same ballpark as the estimates\footnote{In ref.~\cite{Boucaud:2008gn},
 $g^2 \langle A^2\rangle$ is evaluated through an OPE formula with a tree-level Wilson coefficient to be 5.1 GeV$^2$
 at a renormalization point of 10 GeV.} obtained from the SU(3) analysis of the Taylor coupling
 in ref.~\cite{Boucaud:2008gn}. This strongly supports that the OPE analysis indeed captures the kinematical 
 structure for the form factor $H_1$.  
 On the other hand, the infrared mass scale, as supposed, is of the order of 1 GeV
 and the (very close to 1) value for $\widetilde{Z}_1^{-1}$ accounts reasonably for perturbative
 value of $H_1$ that we approximated by a constant. Thus, after paying the economical price of 
 incorporating only these two more parameters, both taking also reasonable values, we are left with 
 a reliable closed formula for the form factor at any gluon or ghost momenta. This is a very 
 useful ingredient for the numerical integration of the GPDSE.

The SU(3) results published by the authors of ref.~\cite{Sternbeck:2005re} appear to be
very noisy and cannot be invoked to properly discriminate whether a constant behaviour close to 1
or \eq{eq:H1model} accounts better for them.
However,  we can  now take the mass parameter, $m_{\rm IR}$, from the
previous SU(2) analysis and the well-known SU(3) value for $g^2 \braket{A^2}$,
assume $\widetilde{Z}_1^{-1}=1$ and use \eq{eq:H1model} to predict
the ghost-gluon transverse form factor, $H_1$. This is shown in the right plot of in Fig.~\ref{fig:lat} for
three different kinematical configurations and, as can be seen, the deviations from 1
appear to be very small in all the cases and compatible with the results shown in
Fig.~4 of ref.~\cite{Sternbeck:2005re} for the vanishing gluon momentum case.


\section{The Taylor kinematics and the asymmetric gluon-ghost vertex}

An especially interesting kinematical configuration in the ghost-gluon vertex is the one where the incoming ghost
momentum goes to zero ($k \to 0$). Let us pay some additional attention to it. This particular kinematical limit
is the one where Taylor's non-renormalization theorem provides us
with a result for the bare vertex, which is at least exact up to all
perturbative\footnote{The same argument of Taylor's perturbative proof still works
if one consider the Landau-gauge ghost-gluon vertex DSE: a vanishing ghost momentum entering in the vertex
implies the contraction of the gluon-momentum transversal projector with the gluon momentum itself for any dressed
diagram. Thus, Taylor's theorem is still in order within the non-perturbative DSE framework. Some
attention have been recently paid to the ghost-gluon vertex DSE~\cite{Schleifenbaum:2004id}.} orders:
\beq \label{taylor}
\Gamma^{abc}_{{\rm bare},\mu}(-q,0;q) \ = \ - g f^{abc} q_\mu \ .
\eeq
At any order of perturbation theory, this implies that $H_1(q,0)+H_2(q,0)=1$. According to \eq{eq:H1model}, one would have
\beq\label{H1OPETaylor}
H_1(q,0) \ = \  \widetilde{Z}_1^{-1} \
\left( 1 +  N_C \frac{g^2 \braket{A^2}}{4 (N_C^2-1)} \ \frac {q^2} {q^4+m_{IR}^4}  \ \right)
\ .
\eeq
An interesting question to investigate is
whether such a genuine non-perturbative correction still survives for the full ghost-gluon
vertex in the Taylor limit, and not only for the transverse form factor $H_1$. Said otherwise,
does eq.\eqref{taylor} maintains its validity upon adding the OPE corrections related to $\VA$ to it, or not?

In order to properly address this question, one does not to have to study the ghost-gluon vertex
for asymptotically large legs' momenta, as done in sec.~\ref{sec:OPEsym}, but to study instead
the asymmetric ghost-gluon vertex
\beq
\widetilde{V}^{abc}_\mu(-q,\varepsilon;q-\varepsilon) &=&
\Gamma^{a'b'c'}_{\mu'}(-q,\varepsilon;q-\varepsilon) \ G^{bb'}_{\mu\mu'}(q-\varepsilon) \ F^{aa'}(q) \
F^{cc'}(\varepsilon)
\nonumber \\
&=&
\int d^4x \ e^{i (q-\varepsilon) \cdot x}
\Braket{\ T\left( \widetilde{c}^c(\varepsilon) A^b_\mu(x) \overline{c}^a(0) \right)}
\eeq
where, when taking the limit $\varepsilon^2 \to 0$,
$\widetilde{c}^a(\varepsilon)$ stands for an incoming vanishing ghost field.
A certain care is needed since we have to deal here with
a ``\emph{soft}'' ghost leg, similarly to the situation in ref.~\cite{DeSoto:2001qx}
when analyzing the three-gluon vertex.
According to this, the key point for the OPE analysis of the asymmetric vertex is to apply
the OPE expansion only to the fields carrying hard momenta and take then advantage of
the operator expressions to compute the appropriate
matrix elements in presence of a ghost field with ``soft'' momentum.
Thus, we can apply to the asymmetric ghost-gluon vertex a procedure similar to the one
outlined in sec.~\ref{sec:OPEsym} and obtain
\beq\label{OPEvertex}
\widetilde{V}^{abc}_\mu(-q,\varepsilon;q-\varepsilon)
&=&
\int d^4x e^{i (q-\varepsilon) \cdot x}
\Braket{\widetilde{c}^c(\varepsilon) \ T\left( A^b_\mu(x) \overline{c}^a(0) \right)}
\nonumber \\
&=&
\left(d_1\right)^{ab}_{\mu c'}(q) \ \Braket{\widetilde{c}^c(\varepsilon)
\ : \overline{c}^{c'}(0) :}
\nonumber \\
&+&
\left(d_3\right)^{ab\mu'\nu'}_{\mu c'a'b'}(q) \
\Braket{\ \widetilde{c}^c(\varepsilon) \
: \overline{c}^{c'}(0) A_{\mu'}^{a'}(0) A_{\nu'}^{b'}(0) :}
\ + \ \cdots
\eeq
Then, if one follows standard SVZ techniques to obtain the perturbative expansion of OPE
Wilson coefficients, we will take the perturbative vacuum for the matrix elements of
\eq{OPEvertex} and obtain:
\beq\label{eq:d1}
\widetilde{V}^{abc}_{{\rm pert}, \mu}(-q,\varepsilon;q-\varepsilon) \ = \
\left(d_1\right)^{ab}_{\mu c'}(q,\varepsilon) \ F^{cc'}(\varepsilon) \ ,
\eeq
where $d_1$ can be straightforwardly identified with the perturbative expansion of the
ghost-gluon Green function, $\widetilde{V}$, with an amputated soft-ghost leg. Beyond
this purely perturbative first contribution, we will take the matrix element of
\eq{OPEvertex} with two external soft gluons in the perturbative vacuum:
\beq\label{SVZvertex}
&&
\int d^4x e^{i (q-\varepsilon) \cdot x} \
\Braket{\widetilde{c}^c(\varepsilon) \widetilde{A}_{\mu''}^{a''}(0) \widetilde{A}_{\nu''}^{b''}(0) \
T\left( A^b_\mu(x) \overline{c}^a(y) \right)}
\nonumber \\
&=&
\left(d_3\right)^{ab\mu'\nu'}_{\mu c'a'b'}(q,\varepsilon) \
\Braket{0 | \ \widetilde{c}^c(\varepsilon) \widetilde{A}_{\mu''}^{a''}(0) \widetilde{A}_{\nu''}^{b''}(0) \
: \overline{c}^{c'}(0) A_{\mu'}^{a'}(0) A_{\nu'}^{b'}(0) :  | 0}
\nonumber \\
&=&
\left(d_3\right)^{ab\mu'\nu'}_{\mu c'a'b'} (q,\varepsilon) \
F^{cc'}(\varepsilon) \ \left( G^{a''a'}_{\mu''\mu'}(0) \ G^{b''b'}_{\nu''\nu'}(0)
+ G^{b''a'}_{\nu''\mu'}(0) \ G^{a''b'}_{\mu''\nu'}(0)\right)\,.
\eeq
On the other hand, when considering a non-perturbative vacuum for the matrix elements of the
local operators in the OPE expansion of \eq{OPEvertex} and
for a vanishing incoming ghost momentum $\varepsilon^2$, we will apply the factorization approximation
(shown to work in ref.~\cite{DeSoto:2001qx}) that follows:
\beq
\Braket{\widetilde{c}^c(\varepsilon) \
: \overline{c}^{c'}(0) A_{\mu'}^{a'}(0) A_{\nu'}^{b'}(0) :}
\ \ &\to& \ \
\Braket{0 | \widetilde{c}^c(\varepsilon)
\  \overline{c}^{c'}(0) | 0 \rangle \ \langle \ : A_{\mu'}^{a'}(0) A_{\nu'}^{b'}(0) :}
\nonumber \\
&=&
F^{cc'}(\varepsilon) \ \frac{\braket{A^2}}{4 (N^2_C-1)}  \ \delta^{a'b'} g_{\mu'\nu'} \ .
\eeq
Thus, \eq{OPEvertex} can be written as follows:
\beq
\widetilde{V}^{abc}_\mu(-q,\varepsilon;q-\varepsilon)  &=&
\widetilde{V}^{abc}_{{\rm pert}, \mu}(-q,\varepsilon;q-\varepsilon)
\nonumber \\
&+& \underbrace{
\left(d_3\right)^{ab\mu'\nu'}_{\mu c'a'b'}(q,\varepsilon)
\ \delta^{a'b'} g_{\mu'\nu'}}_{\displaystyle \widetilde{w}^{abc'}_\mu} \
\frac{\braket{A^2}}{4 (N^2_C-1)}   \ F^{cc'}(\varepsilon)
\ + \ \cdots
\label{eq:d1d3}
\eeq
where, now from \eq{SVZvertex},
\beq\label{OPE4}
\widetilde{w}^{abc'}_\mu(q) \ &=&
\frac 1 2 \frac{\int d^4x e^{i q \cdot (x-y)} \
\Braket{\widetilde{c}^c(\varepsilon) \widetilde{A}_{\mu''}^{a''}(0) \widetilde{A}_{\nu''}^{b''}(0) \
T\left( A^b_\mu(x) \overline{c}^a(y) \right)}}
{ G^{b'b''}_{\mu'\mu''}(0) \ G^{a'a''}_{\nu'\nu''}(0) \ F^{cc'}(\varepsilon) }
\ \delta^{a' b'} g_{\mu' \nu'}. \
\eeq

However, in the Landau gauge, the transverse incoming gluon kills the longitudinal contribution for
the ghost-gluon vertex and only the transverse form factor, $H_1$, survives. One needs then
to compute the proper (1PI) ghost-gluon vertex, which can be formally writen as
\beq
\Gamma^{abc}_\mu(-q,k;q-k)  &=& \frac{V^{a'b'c'}_{\mu'}(-q,k;q-k)}{G^{bb'}_{\mu\mu'}(q-k) F^{aa'}(q) F^{cc'}(k)} \ .
\eeq
It is clear that, because of \eq{eq:d1}, one can factor the vanishing ghost propagator out in the
r.h.s. of \eq{eq:d1d3} and this implies that $\widetilde{w}^{abc'}_\mu$ in \eq{OPE4} directly
provides with the correction for the ghost-gluon vertex with the vanishing ``soft'' ghost leg
explicitly amputated. Thus, we can formally define
\beq
\widetilde{w}^{abc'}_\mu(q) \ = \ \widetilde{v}^{a'b'c'}_{\mu'}(q)  \ F^{a'a}(q) \ G^{b'b}_{\mu'\mu}(q) \ ,
\eeq
such that the proper ghost-gluon vertex can be obtained as
\beq\label{OPEamp}
\Gamma^{abc}_\mu(-q,\varepsilon;q-\varepsilon)  \ = \
\Gamma^{abc}_{{\rm pert}, \mu}(-q,\varepsilon;q-\varepsilon) \ + \
\widetilde{v}^{abc}_\mu \
\frac{\braket{A^2}}{4 (N^2_C-1)} \
\ + \ \cdots
\eeq
where
\beq
\widetilde{v}^{abc}_\mu \ = \ 2 I^{[1]} \ ,
\eeq
the factor $2$ coming from the combinatorics and $I^{[1]}$ being the following diagram (no external legs)
\beq\label{I1vertex}
I^{[1]} \altura{2.5} & = & \ghThreeOneR \nonumber  \\
&=& - \altura{1} \frac{N_C}{2} \ g^3 \
\frac{(q-\varepsilon)_\sigma q_\sigma}{q^2 (q-\varepsilon)^2} \ f^{abc} \
\left( q_\mu \ - \ \frac{q_{\mu_2} (q-\varepsilon)_{\mu_2}}{(q-\varepsilon)^2} \ (q-\varepsilon)_\mu
\right)\,.
\eeq
Then, according to \eq{vertMinko}, one finally
obtains:
\beq\label{OPEamp2}
\Gamma^{abc}_\mu(-q,\varepsilon;q-\varepsilon)  \ = \
- g f^{abc} \left( \altura{0.6} q_\mu \ H_1(q,\varepsilon) \ + \
(q-\varepsilon)_\mu \ H_2(q,\varepsilon) \right)
\ + \ \cdots
\eeq
where
\beq
H_1(q,\varepsilon) &=&  H_1^{\rm pert}(q,\varepsilon) \ + \
N_C \ g^2 \frac{(q-\varepsilon) \cdot q}{q^2 (q-\varepsilon)^2} \
\frac{\braket{A^2}}{4 (N^2_C-1)} \ , \nonumber \\
H_2(q,\varepsilon) &=& H_2^{\rm pert}(q,\varepsilon) \ - \
N_C \ g^2 \ \frac{\left((q-\varepsilon)\cdot q \right)^2}{q^2 (q-\varepsilon)^4} \
\frac{\braket{A^2}}{4 (N^2_C-1)} \ .
\label{H12OPE}
\eeq
Thus, after taking the limit $\varepsilon \to 0$, one will have:
\beq
\Gamma^{abc}_\mu(-q,0;q)  \ = \
- g f^{abc} \ q_\mu \left( \ \altura{0.6} H_1^{\rm pert}(q,0) \ + \ H_2^{\rm pert}(q,0) \ \right)
\ = \ - g f^{abc} \ q_\mu \ .
\eeq
This confirms that no non-perturbative OPE correction survives in
the proper ghost-gluon vertex in the Taylor limit, although both form factors $H_1$ and $H_2$ separately undergo such a kind of correction. It is also
straightforward to see that $H_1(q,0)$ obtained from \eq{H12OPE} would be  the same as the one given
in \eq{H1OPETaylor}, obtained for the general kinematics,
when $q^2 \gg m^2_{\rm IR}$.
Furthermore, provided we introduce in \eq{H12OPE} for $H_1$  the same infrared mass scale, $m_{IR}$, we used
to obtain \eq{H1OPETaylor}
we would again be left with the same result. These observations underline a rewarding agreement of our resuts
although they stem from slightly different hypotheses.

\section{Conclusions}

The momentum behaviour of the ghost-gluon vertex has been studied in the framework of the operator product expansion (OPE).
This approach has already been proved in the literature to be very fruitful in describing the running of two-point gluon and quark Green functions
and of the strong coupling computed in several MOM renormalization schemes. In particular, the coupling derived from
the ghost-gluon vertex renormalized in T-scheme, which can be directly computed from nothing else but the ghost and gluon
propagators, led to a very accurate determination of $\Lambda_{\rm QCD}$ when its running with momenta obtained from
the lattice estimate has been confronted with the OPE prediction, although only after accounting properly for a dimension-two gluon
condensate, $\braket{A^2}$. In this note, we followed the same approach to study the ghost-gluon vertex
and payed special attention to  the transverse form factor for the ghost-gluon vertex, precisely the one that plays a crucial
r\^ole for the truncation and resolution of the GPDSE and that, in this context, is usually approximated by a constant.
In this way we  obtained a genuine non-perturbative OPE correction to the perturbative part for the transverse form
factor. A very simple conjecture was then made to extend the OPE description beyond the momentum range where the SVZ factorization
is supposed to work and we were left with a simple model describing the momentum behaviour of the ghost-gluon form factor.
This model was shown to agree pretty well with a lattice SU(2) computation for several kinematical
configurations of the ghost-gluon vertex, when the value for the gluon condensate is in the same ballpark as the one that is estimated
from the running of the T-scheme coupling in SU(3) lattice gauge theory. This successful comparison with the available
lattice data provides us with strong indications that (i) the OPE framework helps to account for the
kinematical structure of the ghost-gluon vertex at intermediate momentum domain
and (ii) that it helps to cook up a reliable model, providing a good parameterization also for the IR domain,
to be plugged into the GPDSE to reproduce the ghost propagator lattice data. These are the main goals of the
paper.

Finally, we also proved that, in the particular kinematical limit associated with the Taylor theorem and hence by the
T-scheme (a vanishing incoming ghost momentum), the corrections to the longitudinal and transverse form factors cancel against
each other and  that the tree-level result for the ghost-gluon vertex is thus recovered.\\

In future work, we foresee to put to use the vertex \eqref{eq:H1model} in combination with the GPDSE, this in order to
investigate at the quantitative level how this vertex allows to bring the analytical predictions of the GPDSE in closer
agreement with the lattice output for the ghost propagator, given the input of the kernel via the lattice gluon propagator.

\bigskip

{\bf Acknowledgements:}
 J. R-Q acknowledges the Spanish MICINN for the support by the research project FPA2009-10773 and ``Junta de Andalucia''
by P07FQM02962. D.~Dudal is supported by the Research-Foundation Flanders (FWO Vlaanderen). We thank A. Le Yaouanc,
J. Micheli and A.~Cucchieri for inspiring discussions, while
J. R-Q acknowledges A. C. Aguilar and J. Papavassiliou for very valuable indications triggering the work.



\providecommand{\href}[2]{#2}\begingroup\raggedright\endgroup

\end{document}